\documentclass[footinbib,aps,pra,reprint,superscriptaddress,nopacs]{revtex4-1}

\usepackage{amsmath,amssymb,bbm,bm,graphicx}
\usepackage{csquotes}
\usepackage{amsfonts}
\usepackage{verbatim}
\usepackage{braket}   % for bra ket notation
\usepackage{mathtools}

\usepackage[absolute]{textpos}
\usepackage{xcolor}

\begin{document}
\title{Quantum frequency combs and Hong--Ou--Mandel interferometry:\\ the role of spectral phase coherence}

\author{Navin B. Lingaraju}
\email{nlingara@purdue.edu}
\affiliation{School of Electrical and Computer Engineering and Purdue Quantum Science and Engineering Institute, Purdue University, West Lafayette, Indiana 47907, USA}

\author{Hsuan-Hao Lu}
\affiliation{School of Electrical and Computer Engineering and Purdue Quantum Science and Engineering Institute, Purdue University, West Lafayette, Indiana 47907, USA}

\author{Suparna Seshadri}
\affiliation{School of Electrical and Computer Engineering and Purdue Quantum Science and Engineering Institute, Purdue University, West Lafayette, Indiana 47907, USA}

\author{Poolad Imany}
\affiliation{School of Electrical and Computer Engineering and Purdue Quantum Science and Engineering Institute, Purdue University, West Lafayette, Indiana 47907, USA}

\author{Daniel E. Leaird}
\affiliation{School of Electrical and Computer Engineering and Purdue Quantum Science and Engineering Institute, Purdue University, West Lafayette, Indiana 47907, USA}

\author{Joseph M. Lukens}
\affiliation{Quantum Information Science Group, Computational Sciences and Engineering Division, Oak Ridge National Laboratory, Oak Ridge, Tennessee 37831, USA}

\author{Andrew M. Weiner}
\affiliation{School of Electrical and Computer Engineering and Purdue Quantum Science and Engineering Institute, Purdue University, West Lafayette, Indiana 47907, USA}

\date{\today}

\begin{abstract}
The Hong--Ou--Mandel interferometer is a versatile tool for analyzing the joint properties of photon pairs, relying on a truly quantum interference effect between two-photon probability amplitudes. While the theory behind this form of two-photon interferometry is well established, the development of advanced photon sources and exotic two-photon states has highlighted the importance of quantifying precisely what information can and cannot be inferred from features in a Hong--Ou--Mandel interference trace. Here we examine Hong--Ou--Mandel interference with regard to a particular class of states, so-called quantum frequency combs, and place special emphasis on the role spectral phase plays in these measurements. We find that this form of two-photon interferometry is insensitive to the relative phase between different comb line pairs. This is true even when different comb line pairs are mutually coherent at the input of a Hong--Ou--Mandel interferometer, and the fringe patterns display sharp temporal features.  Consequently, Hong--Ou--Mandel interference cannot speak to the presence of high-dimensional frequency-bin entanglement in two-photon quantum frequency combs.
\end{abstract}

\maketitle
\section{Introduction}

Two-photon quantum frequency combs -- biphoton frequency combs (BFCs) for short -- have attracted interest in recent years owing to their scalability to high dimensions~\cite{Kues2019}. These states (see Fig.~\ref{BFC_concept}) are characterized by entanglement over discrete pairs of symmetric (energy-matched) comb lines, or frequency bins, consequently referred to as frequency-bin entanglement~\cite{Ramelow2009,Olislager2010}. In analogy to classical frequency combs, the joint spectrum of these states features comb lines equidistant from one another in the frequency domain. The comb-like nature of the biphoton spectrum is well-suited to quantum information processing in the spectral domain~\cite{Lukens2017}. Recent demonstrations include the realization high-fidelity discrete Fourier transform gates~\cite{Lu2018}, parallel qubit rotations using a quantum frequency processor~\cite{LuHOM}, a coincidence-basis controlled-NOT~\cite{LuCNOT}, and single-photon two-qudit gates~\cite{Imany2019,Reimer2019}. These demonstrations have largely used commercial off-the-shelf telecommunications equipment. Furthermore, BFCs have been generated directly in CMOS-compatible optical microresonators by spontaneous four-wave mixing~\cite{Reimer2016,Jaramillo2017,Kues2017,Imany2018}. By appropriately engineering the dispersion of the microresonator, BFCs can be generated almost anywhere across the telecommunications band. These features, coupled with scalability to high dimensions, show the potential for BFCs in, e.g., the development of practical quantum networks.

An important step to advancing this platform is the development of robust methods to certify entanglement, which usually require projecting two-photon states onto different bases~\cite{James2001}. In the spectral domain this projection requires frequency mixing, which has been implemented, albeit probabilistically, using electro-optic phase modulators (EOMs)~\cite{Kues2017,Imany2018a,Imany2018}. The probabilistic nature of the frequency-mixing operation, coupled with the need for high-speed radio-frequency (RF) electronics, makes this method of certifying frequency-bin entanglement challenging. While, in principle, deterministic frequency-mixing operations can be achieved with an alternating series of EOMs and Fourier transform pulse shapers~\cite{Lu2018, LuHOM}, it comes at the cost of increased system loss due to the introduction of additional components. These limitations become very apparent in the case of high-dimensional frequency-bin entanglement, where the measurement process quickly becomes time-consuming and a major strain on resources.
 
This motivates the need for an alternative method to certify high-dimensional frequency-bin entanglement. It was recently suggested that HOM interference, which can be sensitive to delays on the order of femtoseconds, could be used to probe features in the joint temporal correlation of a BFC~\cite{Xie2015,Chang2019} and, therefore, detect the presence of high-dimensional frequency-bin entanglement. Indeed, one of the interesting features of BFCs, noted in early experiments~\cite{Lu2003}, is their production of HOM ``revivals''; i.e., the initial coincidence dip at zero delay reappears (as dips or peaks) at multiples of half the inverse free spectral range (FSR)~\cite{Lu2003, Zavatta2004, Sagioro2004}. As each dip has a duration set by the total biphoton bandwidth, HOM revivals are certainly a broadband effect and could appear, initially, to be an effect exploiting broadband phase coherence as well. However, in this paper, we demonstrate this is not the case.

Our work builds on a body of literature utilizing HOM interference to characterize frequency-entangled quantum states. For example, Ramelow \emph{et al.}~\cite{Ramelow2009} showed that antibunching at a beam splitter can serve as an indicium of two-dimensional frequency-bin entanglement. Very recently, Jin and Shimizu~\cite{Jin2018}, in drawing parallels between classical and quantum interferometry, related the Fourier transform of an HOM interference trace to a projection of the joint spectral intensity (JSI) along the difference-frequency axis. Both works show that HOM interference carries some information about entanglement and the joint spectrum. However, the former was limited to a two-dimensional frequency-bin state and the latter made several simplifying assumptions on the biphoton state regarding its symmetry and phase. Thus there exists an important need for examining HOM interference in the case of \emph{both} high-dimensional frequency correlations and arbitrary phase relationships. In this article, we derive and experimentally demonstrate complete insensitivity of HOM interference to phase coherence across spectrally distinct pairs in broadband BFCs. Accordingly, HOM interferograms cannot serve as an indicium of high-dimensional frequency-bin entanglement. Our results remain consistent with well-established theory, while simultaneously shedding light on the limitations of HOM interference in characterizing quantum states.

\section{Background}
\label{Background}
\begin{figure}[tb!]
\centering\includegraphics[width=\columnwidth]{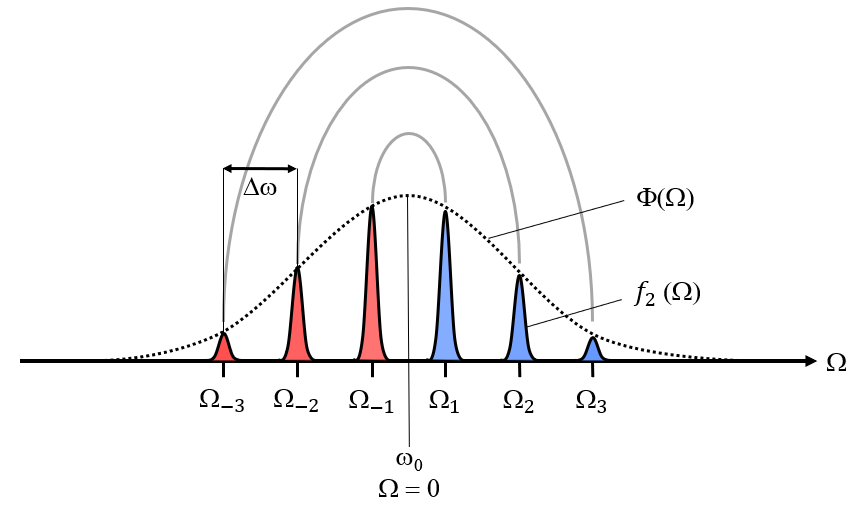}
\caption{General form of BFC frequency bins. The frequency offset ($\Omega$) is defined with respect to $\omega_0$ ($2\omega_0$ = pump laser frequency). Bins centered at $\Omega_{p}$ are carved from a broadband biphoton spectrum $\Phi(\Omega)$, and adjacent bins are separated from one another by $\Delta\omega$. Although the weight of each frequency bin is set by the profile of the broadband biphoton spectrum, all bins are carved from spectral filters with identical lineshapes [$f_{2}(\Omega)$ shown as an example]. Grey lines highlight pairwise entanglement across symmetric (energy-matched) bins.}
\label{BFC_concept}
\end{figure}

In our experiments, we generate entangled photon pairs through spontaneous parametric down-conversion (SPDC) of a continuous-wave (CW) laser at frequency $2\omega_0$ whose linewidth is much narrower than the frequency bins themselves. These frequency bins are centered at the optical frequencies $\omega_0+\Omega_p$, where the baseband offset $\Omega_p$ is defined by:
\begin{equation}
\label{freqDef}
\Omega_p = \begin{cases} \left(p-\frac{1}{2}\right)\Delta\omega & p=1,2,...\\
\left(p+\frac{1}{2}\right)\Delta\omega & p=-1,-2,... ,
\end{cases}
\end{equation}
with $\Delta\omega$ the FSR of the comb modes. In this formulation, we are assuming the degeneracy point of the pump laser lands precisely between the $p = -1$ and $p = 1$ bins. (See Fig.~\ref{BFC_concept} for a graphical depiction of our frequency bin definitions.) In this regime, bins with indices $\pm p$ are frequency-entangled, and we can define a fundamental frequency-bin unit with photon $A$ in bin $p$ and photon $B$ in bin $-p$, according to
\begin{equation}
\label{JML1}
\ket{p,-p}=\int d\Omega \,  \Phi(\Omega) f_p(\Omega)
\ket{+\Omega,-\Omega}_{AB}
\end{equation}
where $\Phi(\Omega)$ specifies the generated broadband two-photon spectral amplitude, $f_p(\Omega) = f(\Omega -\Omega_p)$ is a lineshape function [symmetric about the center $f(-\Omega)=f(\Omega)$, and normalized such that $\int d\Omega |f(\Omega)|^2 =1$], and $\ket{+\Omega,-\Omega}_{AB} \equiv \hat{a}^\dagger(\omega_0+\Omega) \hat{b}^\dagger(\omega_0-\Omega)\ket{\mathrm{vac}}$ describes a pair of photons -- one in path $A$ with frequency $\omega_0+\Omega$ and the other in path $B$ with frequency $\omega_0-\Omega$. Experimentally, the lineshape function is created by a pulse shaper, which defines spectral filters in either or both of the photons' paths such that filters centered at different frequencies are nonoverlapping, i.e., $f_p(\Omega)f_q(\Omega) \approx 0$ $\forall p\neq q$~\cite{Lukens2014, Imany2018a}.

Note that, because of our use of a CW pump, there exists frequency entanglement within the fundamental frequency-bin unit -- state $\ket{p,-p}$. In some contexts, it is valuable to instead produce separable frequency comb lines, in which case the photon spectrum \emph{within a single line} is uncorrelated with its energy-matched partner~\cite{Reimer2016,Kues2017}. Nevertheless, since our tests here consider phase effects across different frequency bins, our general findings on phase coherence extend to separable bins as well, which can be verified through calculations similar to those below. Finally, our choice of Eq.~(\ref{JML1}) as the fundamental frequency-bin unit is motivated by typical experimental measures, such as the JSI. Common in BFC experiments~\cite{Reimer2014,Kues2017,Imany2018,Xie2015,Jaramillo-Villegas2017,LuHOM}, the JSI reveals the spectral correlations shared by two photons, but provides no information on their mutual phase. In this context, the state $\ket{p,-p}$ corresponds to a single BFC line in the joint spectrum, and our fundamental question can then be posed as, how does phase coherence between pairs with different indices $p$ impact HOM interference?

Now, the state expressed in Eq.~(\ref{JML1}) by itself does not lead to interference in an HOM experiment with slow (integrating) detectors because the two photons $A$ and $B$ share no common frequencies. A superposition of two frequency-bin units, though, with flipped frequency correlations does produce a nontrivial interferogram. Consider the input state
\begin{equation}
\label{JML3}
\ket{\psi_p(\alpha_p)} \propto \ket{-p,p}_{AB} + e^{i\alpha_p}\ket{p,-p}_{AB},
\end{equation}
which we call the ``$p^{\mathrm{th}}$ comb line pair'' ($p\in \mathbb{N}$). If we introduce a delay $\tau$ in path $A$ and mix the two spatial modes at a 50:50 beam splitter, the output electric field operators (in modes labeled $C$ and $D$) can be written as~\cite{Mandel1995}
\begin{equation}
\begin{aligned}
\label{Efield}
\hat{E}_C^{(+)}(t) = \frac{1}{\sqrt{2}} \int d\omega_1\,\left[ e^{i\omega_1\tau}\hat{a}(\omega_1) + i\hat{b}(\omega_1) \right] e^{-i\omega_1 t} \\
\hat{E}_D^{(+)}(t) = \frac{1}{\sqrt{2}}\int d\omega_2\,\left[ ie^{i\omega_2 \tau}\hat{a}(\omega_2) + \hat{b}(\omega_2) \right] e^{-i\omega_2 t}.
\end{aligned}
\end{equation}
The total coincidences between the two output ports registered in a time interval $\Delta t$ is then proportional to
\begin{align}
\begin{split}
\label{JML5}
&C_p(\tau;\alpha_p) \propto \\&\int_{\Delta t} dt \int_{T_R} dT\, \left|\left\langle\mathrm{vac} \bigg| \hat{E}_C^{(+)}(t+T) \hat{E}_D^{(+)}(t) \bigg| \psi_p(\alpha_p) \right\rangle  \right|^2,
\end{split}
\end{align}

where the integral over $T$ extends over the detector resolving time $T_R$, assumed much longer than the wavepacket duration. Inserting Eqs.~(\ref{JML3}) and (\ref{Efield}) into Eq.~(\ref{JML5}), and making use of Eqs.~(\ref{freqDef}) and (\ref{JML1}), we arrive at the output coincidence counts
\begin{align}
\begin{split}
\label{JML2}
&C_{p}(\tau;\alpha_p) = \\&K_p \left\{1 - \operatorname{\mathbb{R}e} \left[e^{-i\alpha_p} e^{i(2p-1)\Delta\omega\tau} \int d\Omega\, |f(\Omega)|^2 e^{-2i\Omega\tau}   \right]  \right\}
\end{split}
\end{align}
 
where $K_p$ is a weight specifying the relative probability of populating the comb line pair $\ket{\psi_p(\alpha_p)}$. To arrive at this form, we have assumed that the original spectrum $\Phi(\Omega)$ has symmetric probability about the degeneracy point [$|\Phi(-\Omega)|=|\Phi(\Omega)|$]; that its amplitude variation within a bin is negligible, so that $|\Phi(\Omega)| f_p(\Omega) \approx |\Phi(\Omega_p)| f_p(\Omega)$; and that the phase-mismatch is dominated by terms linear and quadratic in offset frequency $\Omega$. All these are satisfied by our degenerate type-II source, and could be adapted with slight modifications to any of the BFC platforms previously demonstrated. We define the delay so that $\tau=0$ corresponds to the center of the dip when $\alpha_p=0$.

The form of Eq.~(\ref{JML2}) offers insight into the basic features of HOM interference for the comb line pair $\ket{\psi_p(\alpha_{p})}$. The trace is characterized by a fringe pattern that oscillates at frequency $(2p-1)\Delta\omega$, the spacing between the two frequency bins in the $p^{\mathrm{th}}$ pair. This fringe pattern is bounded by an envelope function that is related, by a Fourier transform, to the spectral lineshape common to all frequency bins. The phase offset between the oscillation pattern and the envelope depends on $\alpha_{p}$, i.e., the phase between $\ket{-p,p}_{AB}$ and $\ket{p,-p}_{AB}$; the two-photon basis states that form the $p^{\mathrm{th}}$ comb line pair.  

Finally, if we then consider a high-dimensional frequency-bin superposition state of the form
\begin{equation}
\label{super}
\ket{\Psi} = \sum_{p\in \mathbb{N}} c_p \ket{\psi_p(\alpha_p)},
\end{equation}
a similar calculation leads to an HOM interference fringe pattern given by
\begin{equation}
\label{takeThat}
C_{\ket{\Psi}}(\tau) = \sum_{p\in \mathbb{N}} |c_p|^2 C_p(\tau; \alpha_p),
\end{equation}
with $C_p(\tau; \alpha_p)$ as defined in Eq.~(\ref{JML2}). The phases of the  coefficients $c_p$ -- i.e., any defined phase relationships between different comb lines pairs -- do not appear in the expression for the fringe pattern. Accordingly, calculating the HOM trace for the completely mixed state with density matrix
\begin{equation}
\label{mixed}
\hat{\rho} = \sum_{p\in \mathbb{N}} |c_p|^2 \ket{\psi_p(\alpha_p)}\bra{\psi_p(\alpha_p)}
\end{equation}
returns an interferogram that is \emph{identical} to that of the pure state case, i.e., $C_{\hat{\rho}} (\tau) = C_{\ket{\Psi}} (\tau)$, since the coincidence probability of a given density matrix $\hat{\rho}$ is simply the weighted sum of the probability of each constituent state $\ket{\psi_p(\alpha_p)}$. In other words, HOM interference cannot distinguish between a true high-dimensional frequency-bin-entangled state and an incoherent mixture of comb line pairs. Clarifying this point is the main contribution of this paper, and in the following we confirm this relationship directly in BFC experiments.

\section{Results}
\label{Results}

\begin{figure}[b!]
\centering\includegraphics[width=\columnwidth]{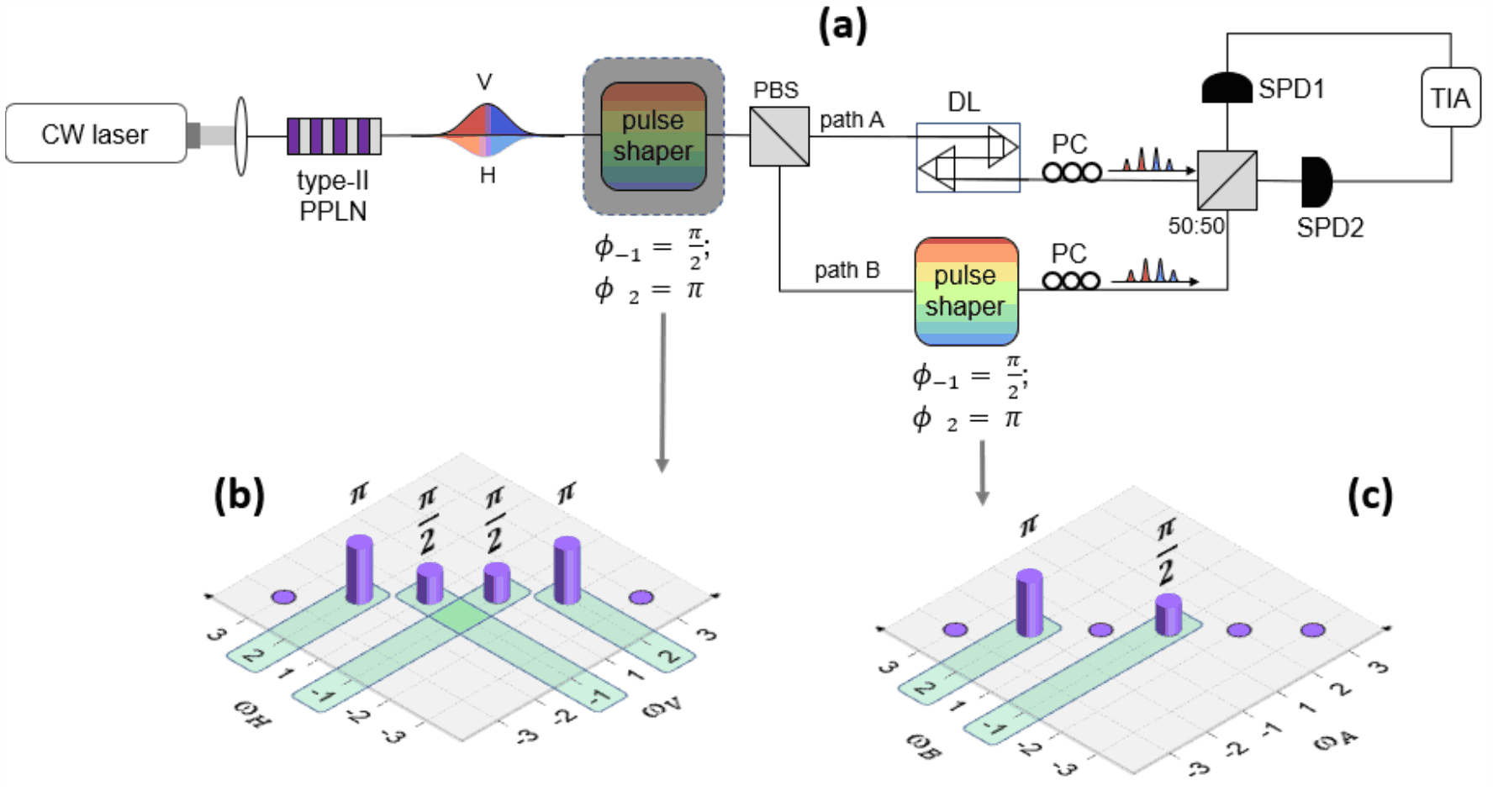}
\caption{(a) Experimental arrangement (see text for details). PPLN, periodically-poled lithium niobate waveguide (HCPhotonics). Pulse shaper (Finisar). PBS, fiber-based polarizing beam splitter. PC, polarization controller. 50:50, fiber-based 50:50 beam splitter. SPD, single-photon detectors (Quantum Opus). TIA, time interval analyzer (PicoQuant). (b-c) Joint phase accumulated by a 6-bin BFC when phases of $\frac{\pi}{2}$ and $\pi$ are applied to frequency bins $-1$ and $2$, respectively. (b) For the arrangement with the pulse shaper placed before the PBS, any applied phase is common to both photons. (c) This is not the case when the pulse shaper is placed within the HOM interferometer as now one of the photons can accumulate phase distinct from the other.}
\label{setup}
\end{figure}

Figure~\ref{setup}(a) shows our experimental setup. We generated photon pairs through collinear, type-II SPDC in a fiber-coupled, periodically poled lithium niobate (PPLN) ridge waveguide with a quantum efficiency on the order of $10^{-7}$. The PPLN was pumped with CW laser light at a wavelength of 779.40~nm (frequency $2\omega_0 = 2\pi\times 384.6$~THz), chosen to ensure that time-energy entangled photons generated by the down-conversion process were fully degenerate. Owing to type-II phase matching in the crystal, the down-converted photons were separated deterministically with a $1\times2$ fiber-based polarizing beam splitter (PBS). Longpass filters, in conjunction with a pair of collimators (not shown), were used in both arms of the HOM interferometer to reject residual pump light.

A Fourier-transform pulse shaper~\cite{Weiner2000, Weiner2011} was used to carve biphoton frequency combs from the continuous down-conversion spectrum -- i.e., to produce the lineshape functions $f_p(\Omega)$ in Eq.~(\ref{JML1}) -- as well as to apply phases to the comb lines. In one set of experiments, the results of which are presented in Appendix~A, the pulse shaper was placed before the PBS (shaded pulse shaper in Fig.~\ref{setup}). Because HOM interference is intrinsically insensitive to any common phase experienced by the two photons~\cite{Steinberg1992,Steinberg1992A,Abouraddy2002}, the HOM traces remained unchanged, regardless of the applied spectral phase modulation, even if fluctuating over the course of a measurement. By contrast, when the pulse shaper is placed in one of the arms of the HOM interferometer (path $B$), it is possible to manipulate the HOM interference pattern in a variety of ways. The results discussed in this section pertain to this arrangement. 

The other arm of the HOM interferometer (path $A$) included two optical delay lines: a 330~ps manually actuated delay line and a 167~ps motorized delay line. The former was used for coarsely matching the delay between path $A$ and $B$, while the latter for scanning the path length difference in two-photon interference experiments. Photons $A$ and $B$ were mixed at a fiber-based 50:50 beam splitter. Interference between two-photon probability amplitudes at the 50:50 beam splitter was observed by monitoring two-photon coincidences between the output ports of the HOM interferometer using superconducting nanowire single-photon detectors  and a time interval analyzer. An important point to note here is that with the pulse shaper in path $B$ alone, it can act on just one of the photons. However, this is sufficient to generate comb-like correlations in the joint spectrum since coincidence measurements postselect energy-matched frequencies of the photon in path $A$~\cite{Lukens2014}.

Reported coincidences correspond to the total events logged in a single 256~ps-wide histogram bin centered on the point of zero path length difference in the interferometer -- an interval larger than the combined timing jitter of our single-photon detectors ($\sim$110~ps), the duration of the filtered biphoton wavepacket ($\sim$60~ps), and the total delay span, thus satisfying the integrating detector limit. Any error bars included with experimental data give the standard deviation of the counts assuming Poissonian statistics. Solid curves in grey correspond to results predicted by theory, while solid lines in other colors simply connect data points for the sake of visual representation.

\subsection{Characterization of down-conversion spectrum and HOM interferometer}

\begin{figure}[bt!]
\centering\includegraphics[width=\columnwidth]{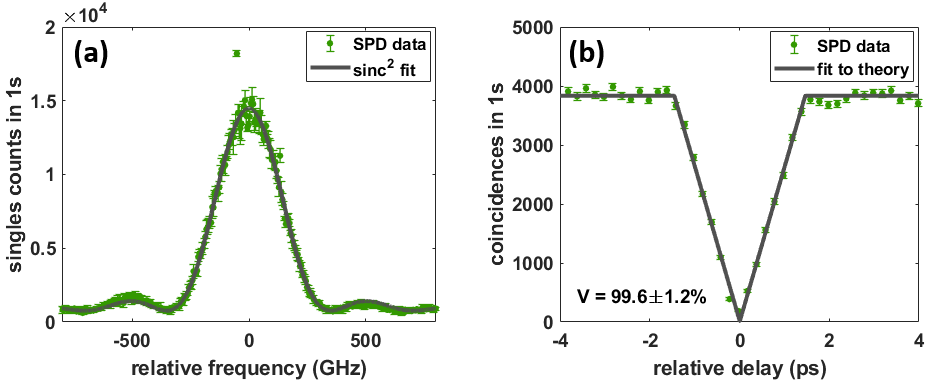}
\caption{(a) Spectrum of the photon in path $B$, acquired by scanning an 18~GHz filter with the pulse shaper. (b) HOM interference trace of the unfiltered SPDC spectrum. A visibility of $(99.6 \pm 1.2)\%$ was obtained without background subtraction.}
\label{SPDC_HOM}
\end{figure}

To measure the unfiltered photon spectrum, we scanned an 18~GHz passband on the pulse shaper and recorded single photon counts for each filter setting, with the results shown in Fig.~\ref{SPDC_HOM}(a). The marginal single-photon spectrum has a characteristic sinc-squared profile with a full-width at half-maximum (FWHM) of $\sim$310~GHz. Subsequently, the path length difference in the HOM interferometer was scanned to obtain the interferogram for the unfiltered biphoton. Results from coincidence measurements are presented in Fig.~\ref{SPDC_HOM}(b) without background subtraction, showing excellent agreement with the triangular shape expected from theory. The high visibility of $(99.6\pm1.2)\%$ is indicative of both the stability of the interferometer, as well as the spectral indistinguishability of the two down-converted photons.   

\subsection{Sensitivity to phase between comb line pairs}
\begin{figure}[bt!]
\centering\includegraphics[width=\columnwidth]{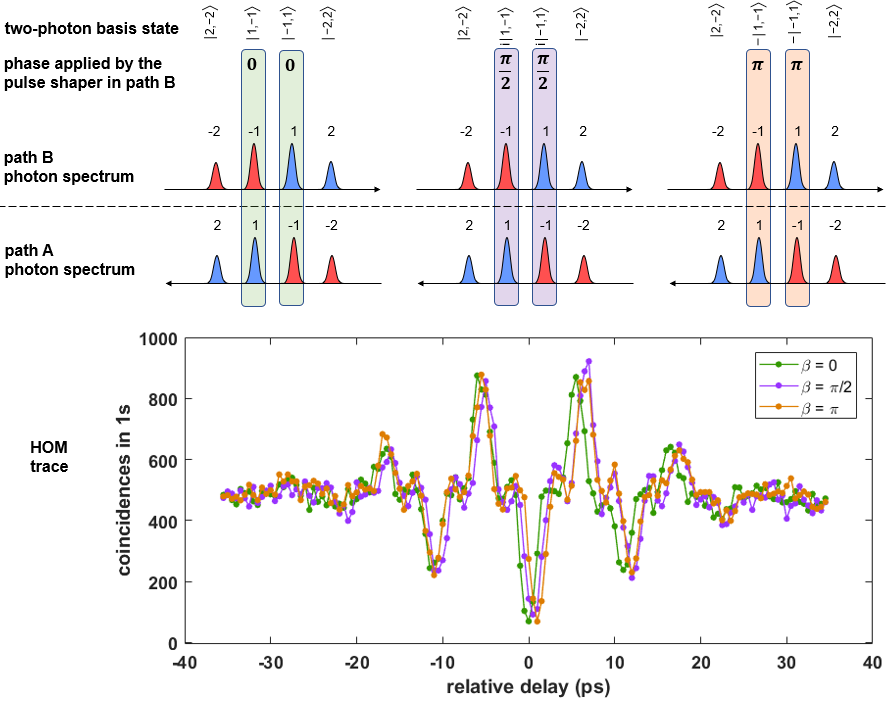}
\caption{HOM interference traces for $\ket{\Psi(0)}$, $\ket{\Psi(\frac{\pi}{2})}$, and $\ket{\Psi(\pi)}$, i.e., three states with different values of $\beta$ -- the relative phase between frequency bin pairs $\ket{\psi_1(0)}$ and $\ket{\psi_2(0)}$. In this and subsequent figures, the axes of the photon spectra in path $A$ and path $B$ are reversed with respect to one another. This has been done to align frequency bins that contribute to the same two-photon basis state. For clarity, high and low frequencies are colored blue and red, respectively, and are defined with respect to the center of the photon spectrum. To facilitate an easy comparison between overlaid HOM interference traces, experimental data (solid circles) for consecutive delay steps are connected by straight lines.}
\label{phase_between_together}
\label{vary_phase_between}
\end{figure}

A typical frequency-degenerate BFC will contain a superposition of many frequency bin pairs in the form of Eq.~(\ref{JML3}). In our experiments, we worked with a simple, but illustrative, system -- a four-bin BFC, or two comb line pairs. This high-dimensional system utilized four frequency bins, centered at offset frequencies $\Omega_{-2}$, $\Omega_{-1}$, $\Omega_1$, and $\Omega_2$, all carved by the pulse shaper. In our experiments, the two down-converted photons are separated based on their polarization state. Consequently, either photon can populate any of the four comb lines, leading to four-dimensional single-photon subspaces. This is unlike BFCs with copolarized photons that can only be distinguished based on their frequency~\cite{Reimer2016, Jaramillo-Villegas2017, Kues2017, Imany2018}, in which case the signal (idler) photon populates only the upper (lower) half of the spectrum. Each comb line was given a Gaussian profile with an intensity FWHM of 17~GHz; the FSR was chosen to be 90~GHz. This effective fill factor ($17/90\approx 0.2$) offered an acceptable trade-off between the overall count rate and the number of revivals in the HOM interference trace. Introducing a phase difference $\beta$ between the two pairs, the four-bin quantum state becomes
\begin{equation}
\label{JML4}
\ket{\Psi(\beta)} \propto e^{i\beta}\ket{\psi_1(0)}+\ket{\psi_2(0)},
\end{equation}
with $\ket{\psi_p(\alpha_p)}$ as defined in Eq.~(\ref{JML3}), and the phase within each comb line pair $\alpha_p$ set to zero. The phase difference $\beta$ can be introduced by applying phase $\beta$ to both the $-1$ and $1$ frequency bins of $\psi_1(0)$, thus keeping the phase within the pair fixed at $\alpha_1 = 0$. To examine the sensitivity of the HOM interferometer to $\beta$, we recorded HOM interferograms for three cases: $\beta\in\{0,\pi/2$, $\pi\}$. An illustration of these operations, the ensuing modification of two-photon basis states, and the corresponding HOM interference traces are shown in Fig.~\ref{vary_phase_between}. Note that these interferograms exhibit antibunching in contrast to those presented in prior work~\cite{Lu2003,Xie2015}. (See Fig.~\ref{PSinFront} in the Appendix for more details.)

Coincidences were counted over 1~sec intervals and are reported without background subtraction. The visibilities of these traces are lower than that of the unfiltered down-conversion spectrum [Fig.~\ref{SPDC_HOM}(b)] as we increased the pair production rate to counteract the significant reduction in counts associated with carving four discrete bins from a broadband spectrum. Although there is some noticeable drift in the interferometer that occurred during acquisition of the $\beta = 0$ interference trace, there is no discernable difference between the features in HOM interference for different values of $\beta$.

Having shown that HOM interference is insensitive to the relative phase between comb line pairs in a coherent superposition state, we take the next step and examine whether the HOM interferometer can, in any way, distinguish between a coherent superposition state and the corresponding mixed state. In other words, can one distinguish between \emph{high-dimensional} and \emph{two-dimensional} frequency-bin entanglement via an HOM measurement? 

\begin{figure}[tb!]
\centering\includegraphics[width=\columnwidth]{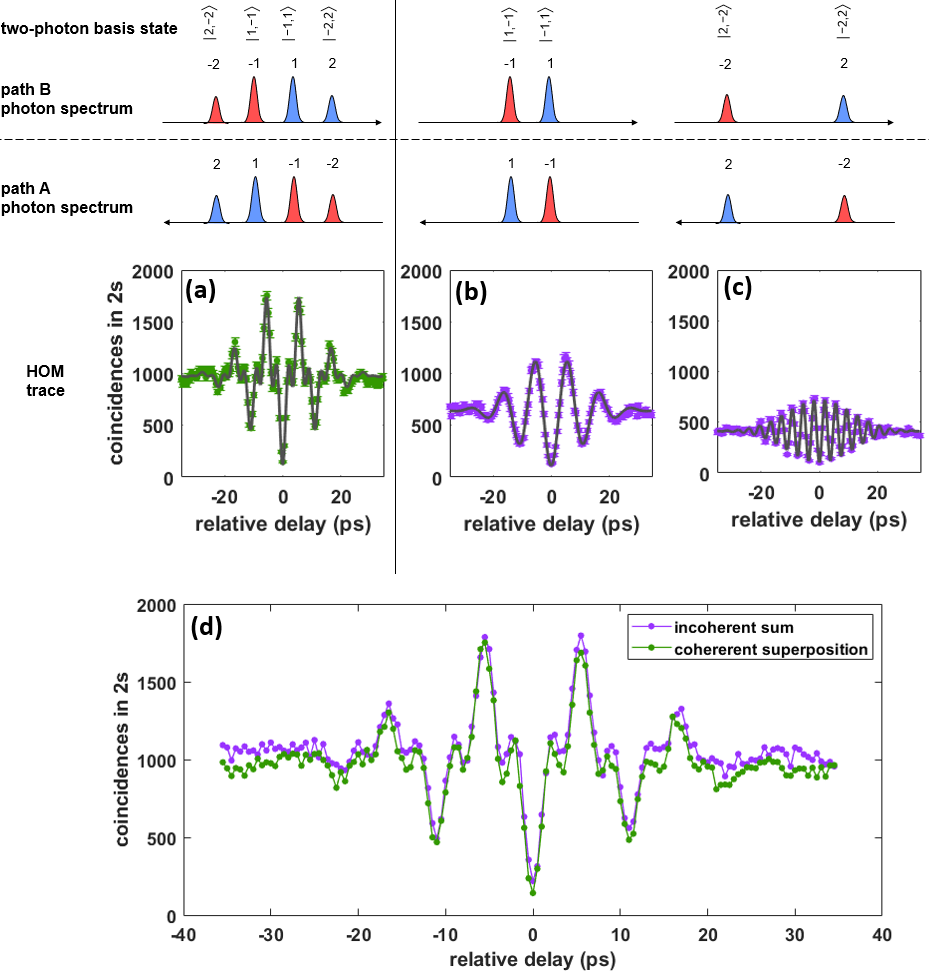}
\caption{HOM interference traces for (a) a coherent superposition state -- $\ket{\Psi(0)}$ = $\ket{\psi_1(0)}$ + $\ket{\psi_2(0)}$, (b) comb line pair $\ket{\psi_1(0)}$ and (c) comb line pair $\ket{\psi_2(0)}$. (d) A comparison between HOM interference for a coherent superposition state (trace (a)) and a mixture of the constituent comb line pairs (trace (b) + trace (c)).}
\label{pure_pair1_pair2}
\end{figure}

For this comparison, we recorded an HOM interferogram for the state $\ket{\Psi(0)}= \ket{\psi_1(0)}+\ket{\psi_2(0)}$ [Fig.~\ref{pure_pair1_pair2}(a)] -- a coherent superposition of two comb line pairs. For the corresponding mixed state [cf. Eq.~(\ref{mixed})], there is no stable phase relationship between the constituent comb line pairs. In other words, the mixture is simply an incoherent sum of two pure states -- $\ket{\psi_1(0)}$ and $\ket{\psi_2(0)}$. Consequently, an HOM interference trace for the mixture can be constructed by recording interferograms for $\ket{\psi_1(0)}$ and $\ket{\psi_2(0)}$ individually [Fig.~\ref{pure_pair1_pair2}(b) and (c)] and adding the two traces together. This is the situation which would result from repeated random emission of either $\ket{\psi_1(0)}$ or $\ket{\psi_2(0)}$, but not both in superposition. The HOM traces for these two cases -- a coherent superposition of two comb line pairs and the corresponding mixture -- are presented on top of one another in  Fig.~\ref{pure_pair1_pair2}(d). Both traces closely track one another, as predicted by the theoretical treatment outlined in Sec.~\ref{Background}. The results in Figs.~\ref{vary_phase_between} and \ref{pure_pair1_pair2} thereby confirm two key points: neither the specific phase between two comb line pairs in a superposition state, nor even the presence of inter-pair coherence more generally, has any bearing on the HOM intereferogram. Therefore, HOM inteference cannot be used to detect or certify high-dimensional frequency-bin entanglement in a BFC.

\subsection{Sensitivity to phase within comb line pairs}
The results of the previous section confirm the inability of HOM interferometry to provide any indication of high-dimensional BFC entanglement. Nevertheless, the results of an HOM diagnostic are not completely independent of frequency-bin entanglement either; for example, Ref.~\cite{Ramelow2009} utilized the fringe pattern from HOM interferometry to estimate off-diagonal elements of a biphoton frequency-bin density matrix, under reasonable assumptions regarding the matrix's form. More generally, prior work~\cite{Wang2006,Eckstein2008,Fedrizzi2009} has established that photon antibunching in an HOM measurement is a sufficient, though not necessary, signature of entanglement; no separable state at the input of an HOM interferometer can produce a coincidence probability greater than $\frac{1}{2}$.

\begin{figure}[tb!]
\centering\includegraphics[width=\columnwidth]{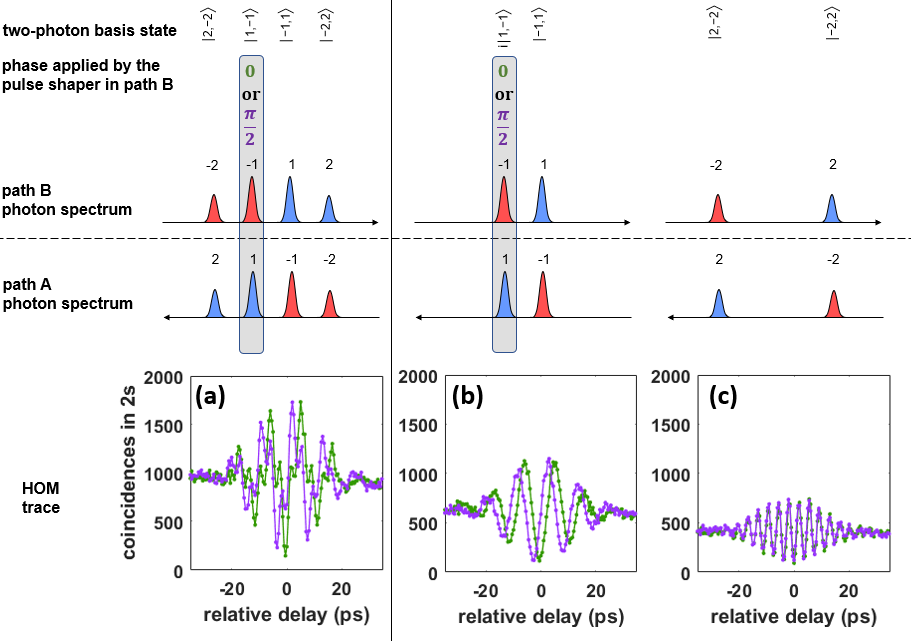}
\caption{HOM interference traces of $\ket{\Psi(0)} = \ket{\psi_1(\alpha_1)} + \ket{\psi_2(0)}$ are shown for two different values of $\alpha_1$, the phase between two-photon basis states $\ket{-1,1}$ and $\ket{-1,1}$. The interferograms in (a) correspond to traces for $\ket{\Psi(0)} = \ket{\psi_1(0)} + \ket{\psi_2(0)}$ (green) and $\ket{\Psi(0)} = \ket{\psi_1(\frac{\pi}{2})} + \ket{\psi_2(0)}$ (purple). HOM interference traces for individual comb line pairs $\ket{\psi_1(\alpha_1)}$ and $\ket{\psi_2(0)}$ are shown in (b) and (c), respectively. Experimental data (solid circles) for consecutive delay steps in each trace connected by straight lines.}
\label{bling}
\end{figure}

Such findings can be shown to be thoroughly consistent with our results, though, by noting a fundamental distinction in BFCs between \emph{two-dimensional} ($d$$=$$2$) and \emph{high-dimensional} ($d$$>$$2$) frequency-bin entanglement. As Eq.~(\ref{JML2}) illustrates, the coincidence rate in an HOM interferometer is sensitive to a phase difference within a comb line pair; i.e., it does depend on the phase $\alpha_p$ between the two terms in Eq.~(\ref{JML3}). To show how $\alpha_p$ affects HOM interference, we repeated the experiments presented in Fig.~\ref{pure_pair1_pair2}, but with one change: phases of $0$ and $\frac{\pi}{2}$ were instead applied to just the $-1$ frequency bin of photon $B$. An illustration of these operations and the corresponding HOM interference traces are presented in Fig.~\ref{bling}.

Figure~\ref{bling}(a) shows interferograms for two different coherent superposition states ($\alpha_1 \in \left\{ 0, \frac{\pi}{2} \right\}$), and Figs.~\ref{bling}(b-c) show fringe patterns for the corresponding comb line pairs $\ket{\psi_1(\alpha_1)}$ and $\ket{\psi_2(0)}$, respectively. As Fig.~\ref{pure_pair1_pair2} demonstrated, the HOM interference trace of a BFC is simply an incoherent sum of the traces of individual comb line pairs. Viewed in this light, the phase-dependent difference between the traces in Fig.~\ref{bling}(a) admits a straightforward explanation: as the phase $\alpha_1$ is varied, all that changes is the contribution from the $\ket{\psi_1(\alpha_1)}$ comb line pair to the overall interferogram. Consequently, the antibunching observed here, as well as across the results in Fig.~\ref{vary_phase_between} and Fig.~\ref{pure_pair1_pair2}, serves only as an indicium of two-dimensional entanglement and nothing more.

\section{Discussion}

The theory and results presented in this article focus on HOM interference as it relates to an emerging platform -- biphoton frequency combs. These states are a subset of time-energy entangled photons and are characterized by pairwise entanglement across energy-matched comb line pairs, as well as a stable biphoton phase relationship between these different pairs. In analogy to classical frequency combs, their joint temporal correlation function possesses fast substructure similar to a mode-locked laser pulse. However, this correlation often cannot be measured directly because the timing resolution (jitter) of single-photon detectors is inadequate to resolve these fast temporal features. To address the question of whether HOM interference can serve as an \emph{indirect} probe of the time correlation function, we examined its sensitivity to changes in the spectral phase of a BFC, finding that HOM interference cannot discern the phase relationship between different comb line pairs in a BFC. 

Intuitively, this can be understood by considering the distinguishability of the various quantum probabilities. Measurement results corresponding to two, spectrally distinct comb line pairs are, in principle, distinguishable for detectors that integrate over many cycles of the frequency separation between the two pairs. Therefore, even if different comb lines are mutually coherent at the input of an HOM interferometer, the only interference detected at the output is that between spectrally indistinguishable comb line pairs. The contributions from different comb line pairs add incoherently, which makes it impossible to distinguish between a coherent superposition state and the corresponding mixed state. It is important to note, however, that these findings depend crucially on operation in the slow-detector regime, so that the electronic resolution integrates over the full biphoton wavepacket [Eq.~(\ref{JML5})]. Given the generally broadband nature of SPDC ($\sim$THz) and typical detector jitters ($\gtrsim$100~ps), this slow-detector case forms by far the most common situation in HOM interferometry. Nevertheless, with sufficiently narrowband photons (e.g., from an atomic source) and correspondingly fast detectors, novel HOM interference phenomena do emerge, such as beating in the coincidence rate for two input photons whose frequencies do not match~\cite{Legero2004, Specht2009}. In this regime, broadband spectral phase has a significant impact, as the detectors can now track the underlying fast temporal fluctuations of the biphoton state. 

On the other hand, even for slow detectors, HOM interference remains sensitive to phase within a comb line pair, because the outcomes leading to a particular coincidence event possess the same frequency content. For even though a red photon at one detector and blue at the other have different frequencies, it is impossible to identify whether one is observing the red (blue) comb-line content of photon $A$ ($B$), or vice versa; a phase difference between these indistinguishable outcomes is precisely what gives rise to HOM interference, and can be exploited for Bell state analysis~\cite{Ramelow2009,Michler1996}. To appropriate Dirac's famous statement, then, we argue, ``In slow-detector HOM, each \emph{frequency pair} then interferes only with itself. Interference between two different \emph{frequency pairs} never occurs.''

Our findings are consistent with prior work on HOM theory~\cite{Mandel1995,GriceWamsley,Agata2017}. For example, one can couch our distinction of the phase \emph{between} and \emph{within} comb line pairs in terms of odd- and even-order polynomial spectral phase. For a degenerate, frequency-entangled pair of photons, HOM is automatically insensitive to any even-order phase experienced by either photon~\cite{Steinberg1992,Steinberg1992A,Abouraddy2002}, whereas odd-order phase on one of the two photons does impact the coincidence probability, with delay (first-order spectral phase) being a clear example. In the BFC case, phase between comb line pairs is that which can always be factored outside of a given pair [e.g., $\beta$ in Eq.~(\ref{JML4})], and so represents an even function of $\Omega$, applying as it does to two energy-matched bins. In contrast, a phase difference within a pair [$\alpha_p$ in Eq.~(\ref{JML3})] produces spectrally asymmetric contributions to the overall phase and thereby is capable of modulating the HOM trace. Thus our observation of HOM's imperviousness to inter-pair phase and sensitivity to intra-pair phase can be viewed as an adaptation of this even/odd-order phase distinction.

\section{Conclusion}
When making their way through an HOM interferometer, spectrally distinct frequency pairs act like silent ships passing in the night, oblivious to the presence of the other. As a result, this form of two-photon interferometry is insensitive to the relative phase between comb line pairs in a biphoton frequency comb and, therefore, cannot be used to detect or quantify high-dimensional frequency-bin entanglement. Our results confirm this, through direct comparison of HOM traces under excitation by phase-coherent and phase-incoherent BFC comb line pairs. However, as was emphasized recently, HOM interference does carry information about the joint spectral intensity (JSI) of the biphoton wavepacket~\cite{Jin2018} and can serve as a substitute in instances where a direct JSI measurement is especially challenging or impractical.  Accordingly, HOM intereferometry proves to be an insightful, powerful, and ubiquitous tool in quantum optics, yet it must never be mistaken for the final word on biphoton phase coherence.

\section*{Appendix A: Pulse shaper in front}
\label{A}

In Fig.~\ref{setup} and Sec.~\ref{Results}, we noted that the pulse shaper used to carve our BFC could be placed just prior to the HOM interoferometer, instead of inside it. In this configuration, the pulse shaper can act on both photons simultaneously, but not independent of one another. In other words, for a given frequency pair, both photons experience the same attenuation and phase shift. Although this limits the form of the joint phase applied by the pulse shaper to the biphoton, one can still examine important aspects of HOM interferometry's sensitivity to high-dimensional frequency-bin entanglement. 

\begin{figure}[b!]
\centering\includegraphics[width=\columnwidth]{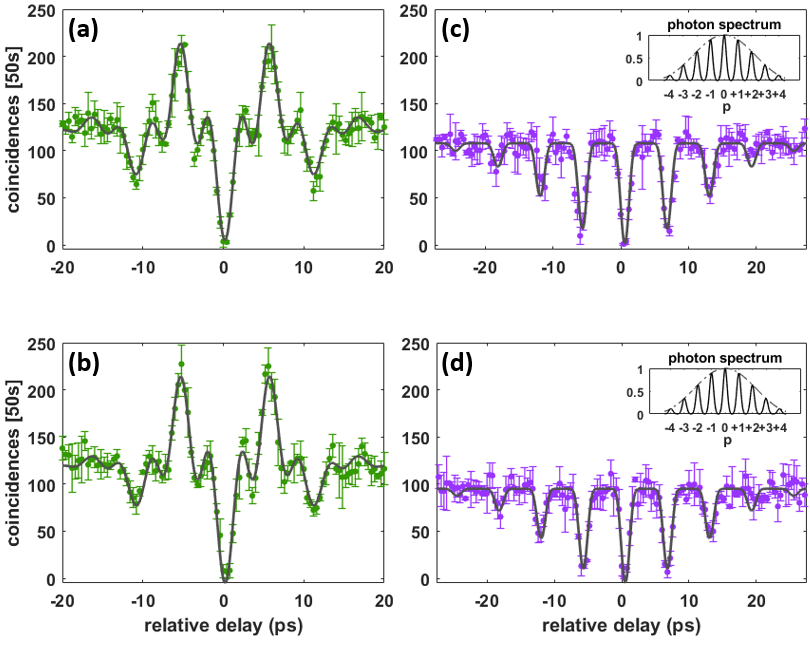}
\caption{HOM interference traces recorded for different biphoton states using the arrangement corresponding to Fig.~\ref{setup}(b), i.e., where the pulse shaper is placed before the HOM interferometer. (a) $\ket{\Psi_a} = \sum_{p=1}^{2} \ket{\psi_p(0)}$, a coherent superposition of four frequency bins. (c) A coherent superposition of nine frequency bins, including a central bin at $p = 0$ (\textit{see inset for the photon spectrum}). The states in (b) and (d) have joint spectral intensities identical to those in (a) and (c), respectively. However, in (b) and (d) the phase relationship between comb line pairs fluctuates within the integration time for each delay step.}
\label{PSinFront}
\end{figure}

Here in Fig.~\ref{PSinFront} we present results acquired using this arrangement. In particular, we compare our base system, (a) a 4-bin BFC with no additional phase applied by the pulse shaper to (b) a 4-bin BFC where the phase relationship between comb line pairs $\ket{\psi_1}$ and $\ket{\psi_2}$ fluctuated over the course of the measurement. The HOM interference traces in (c) and (d) are similarly related to one another but correspond to states with nine frequency bins, including a central bin at $p = 0$. 

The HOM interference trace for our initial state is shown in Fig.~\ref{PSinFront}(a). While this state has the same FSR (90~GHz) as the 4-bin BFC used in Sec.~\ref{Results}, the intensity FWHM of the frequency bins here is 31~GHz. This is apparent from a quick glance at Fig.~\ref{PSinFront}(a); our fringe pattern is now bounded by an envelope that drops off more rapidly than the corresponding trace in Fig.~\ref{vary_phase_between}. Another difference is that the data presented in this section were acquired using InGaAs single-photon detectors. The low efficiency of these detectors ($\sim$10$\%$ and $20\%$), coupled with a gated detection scheme (20~ns window at 1.25~MHz), resulted in a poor coincidence-to-accidental ratio ($\sim$2:1) as compared to the results presented in section~\ref{Results}. Therefore, the coincidence rates in Fig.~\ref{PSinFront} are reported after background subtraction.

Figure~\ref{PSinFront}(b) shows the interferogram for another 4-bin BFC with the same joint spectrum. Unlike the previous case, here the pulse shaper updated the phase applied to the BFC nine different times during the integration window for a single delay step. In particular, three different spectral phase functions were applied three times each: (i)~$\phi(\Omega) = \frac{\phi^{(2)}}{2}\Omega^2$ with $\phi^{(2)}=3.3$~ps$^2$, (ii)~$\phi(\Omega) = \frac{\phi^{(3)}}{6}\Omega^3$ with $\phi^{(3)}=5.1$~ps$^3$, and (iii) a phase profile where each frequency bin received a phase from $\left\{0, \frac{\pi}{12}, \frac{\pi}{6},...,2\pi\right\}$ chosen at random. The spectral phase functions in (i) and (ii) are continuous functions of the angular offset frequency $\Omega$ and varied accordingly across the entire biphoton spectrum, including within each frequency bin. While $\phi^{(2)}$ and $\phi^{(3)}$ remained constant across measurements, the random phase profile was updated with a new set of values every time the function was called. The net effect of these operations was that the joint phase of the BFC included a time-varying contribution from the pulse shaper and ensured that the phase relationship between comb line pairs $\ket{\psi_1}$ and $\ket{\psi_2}$ fluctuated over the course of the HOM measurement. However, these fluctuations in the phase between comb line pairs did not manifest in the interferogram for the state.

Finally, we present a comparison similar to that between (a) and (b), but for the case of a nine-bin state where the degeneracy point $\omega_0$ is shifted to align with one of the comb lines directly, producing a central bin with no matched partner. The marginal spectrum for the photons in this state is shown in the insets of (c) and (d). The FSR of this comb and the width of each frequency bin are the same the states in (a) and (b). As was the case for the previous set of interference traces, (c) corresponds to the case where no additional phase was applied by the pulse shaper, while (d) corresponds to the case where the spectral phase function was updated in the manner described in the previous paragraph. From a comparison of the interferograms in (c) and (d) it is clear that fluctuations in the phase relationship between comb line pairs plays no role in HOM interference. 

Our purpose in highlighting this state is to show the effect of having a frequency bin located at $\omega_0$. Although every comb line pair in this state, except for the central bin, generates a fringe pattern with strong anti-bunching, their results now add in way that obscures the presence of even two-dimensional frequency-bin entanglement.

In summary, these results show that either location of the pulse shaper permits a range of operations useful for examining the sensitivity of HOM interference to phase coherence in biphoton frequency combs.

\section*{Funding}
Laboratory Directed Research and Development Program of Oak Ridge National Laboratory (ORNL); National Science Foundation (NSF) (1839191-ECCS); U.S. Department of Energy, Office of Advanced Scientific Computing Research (Early Career Research Program).

\section*{Acknowledgment}
Some preliminary results for this article were presented at CLEO 2019 as paper number JTu3A.5. A portion of this work was performed at Oak Ridge National Laboratory, operated by UT-Battelle for the U.S. Department
of Energy under contract no. DE-AC05-00OR22725.

\end{document}